\documentclass[]{spie}  

\usepackage{amsmath,amsfonts,amssymb}
\usepackage{graphicx}
\usepackage[colorlinks=true, allcolors=blue]{hyperref}
\usepackage{aas_macros}

\usepackage{mathtools}

\title{Astrometric detection of rocky exoplanets from inexpensive space platforms}

\author{Peter~G.~Tuthill}
\author[a]{George Ayad}
\author[b]{Donald G. Dansereau}
\author[a]{Thomas Holland}
\author[a,b]{Connor J. Langford}
\author[a]{Milo Langker}
\author[c]{Kieran Larkin}
\author[a]{Grace Piroscia}
\author[a]{J. Gordon Robertson}
\author[a]{Karel Valenta}
\author[d]{Tim White}
\author[e]{Asen Christov}
\author[f]{Kyran Grattan}
\author[e]{Sergey Karpov}
\author[g]{Rico Landman}
\author[h]{Pavel Ripka}
\author[i]{Petr Schov\'{a}nek}
\author[j]{Jeffrey Smith}
\author[g]{Frans Snik}
\author[f]{Simon P. Worden}
\author[k]{Ond\u{r}ej Zelenka}

\affil[a]{Sydney Institute for Astronomy, School of Physics, The University of Sydney, NSW 2006, Australia}
\affil[b]{Australian Centre for Robotics, School of Aerospace, Mechanical and Mechatronic Engineering, The University of Sydney, NSW 2006, Australia}
\affil[c]{Nontrivialzeros Research, 22 Mitchell Street, Putney, NSW 2112, Australia}
\affil[d]{Sydney Informatics Hub, Core Research Facilities, University of Sydney, NSW 2006, Australia}
\affil[e]{
Institute of Physics, Czech Academy of Sciences, Na Slovance 2, 182 00 Praha 8, Czechia}
\affil[f]{Breakthrough Prize Foundation, 9 Rue du Laboratoire, L-1911, Luxembourg}
\affil[g]{Universiteit Leiden Sterrewacht, Einsteinweg 55, 2333 CC Leiden, Zuid-Holland, Netherlands}
\affil[h]{MagLab, Faculty of Electrical Engineering, Czech Technical University, Technická 2, 166 27 Praha 6, Czechia}
\affil[i]{Joint Laboratory for Optics, Faculty of Science, Palacký University in Olomouc, 17. listopadu 12, 779 00 Olomouc, Czechia}
\affil[j]{SETI Institute, 339 Bernardo Ave, Suite 200, Mountain View, CA 94043, USA}
\affil[k]{Astronomical Institute, Czech Academy of Sciences, Boční II 1401, 141 00 Praha 4, Czechia}

\authorinfo{e-mail: peter.tuthill@sydney.edu.au}

\begin{document} 
\maketitle

\begin{abstract}
The detection of orbiting exoplanets by way of astrometric monitoring of the host star to reveal gravitationally induced reflex motion requires extreme precision and measurement stability when targeting Earth-mass objects. 
Despite this, astrometry is widely regarded as the most promising avenue for the most desired of all targets: true Earth analogue systems in the solar neighbourhood. 
Formerly thought to require flagship class investment, recently several smaller concepts such as the TOLIMAN and SHERA missions have shown that detections are possible, at least in principle, from more modest space platforms. 
A key enabling feature is that these missions exploit narrow-angle astrometry. 
Observing only multiple star systems where the target provides its
own astrometric reference greatly alleviates the challenges confronted by traditional astrometry with its reliance on distant, faint field stars. This principle underlies the TOLIMAN mission, for which progress and status is reported. However, only a tiny handful of nearby star systems are suited to such a TOLIMAN-like architecture. 
This paper discusses speculative optical concepts featuring ultra-stable telescopes of modest ($<$0.5m) aperture designed for astrometric monitoring of isolated single stars at the micro-arcsecond scales required for the discovery of Earth mass objects orbiting sun-like stars in temperate orbits. 
In simulation, highly sought-after neighbourhood systems such as Epsilon Eridani and Tau Ceti can be exhaustively searched for rocky planets
with dedicated observing campaigns spanning several years, all from relatively inexpensive orbital platforms.

\end{abstract}

\keywords{Exoplanetary detection by Astrometric monitoring; 
High stability space telescopes}

\section{TOLIMAN: Overview}
\label{sec:intro}  

The TOLIMAN mission\cite{2026TOLIMAN_JATIS} addresses one of the most compelling unanswered questions in astronomy and astrobiology: does the nearest stellar system to the Sun, Alpha Centauri AB, host rocky planets in temperate orbits? Despite the remarkable success of exoplanet science over the past three decades, existing detection techniques have made only limited progress towards answering this question.
TOLIMAN adopts a fundamentally different approach based on ultra-precise narrow-angle astrometry. By continuously monitoring the relative separation of the Alpha Centauri AB binary, the mission seeks to detect the minute gravitational perturbations induced by orbiting planets. Such signals manifest as subtle deviations in the binary orbit and require measurement precisions at the microarcsecond level—among the most demanding ever attempted in space astronomy.

Achieving this performance within the constraints of a 16U CubeSat presents formidable scientific and engineering challenges. To meet them, TOLIMAN has developed a novel optical architecture that combines a diffractive pupil element with an exceptionally stable 125\,mm silicon-carbide telescope. The mission's scientific objectives drive requirements for thermal stability, attitude control, and data processing to levels more commonly associated with large flagship-class space observatories, yet these capabilities must be delivered within a tightly constrained mass, volume, and cost envelope.

The TOLIMAN mission was recently described in detail in Tuthill et al. (2026)\cite{2026TOLIMAN_JATIS}, tracing the path from its ambitious scientific goals to the innovative engineering solutions that make them achievable. Particular emphasis is placed on the mission's unique optical system and the spacecraft technologies required to unlock a mode of exoplanet detection inaccessible to current instruments. With most major subsystems now manufactured or under contract, TOLIMAN is progressing towards launch later this decade as discussed in the following sections, demonstrating how a focused, low-cost mission built on novel principles can deliver transformational science.

\section{The TOLIMAN optical system} \label{sec:asgard}

A conventional circular aperture concentrates light into an Airy core that is optimal for centroiding only in an ideal, aberration-free system. TOLIMAN instead uses a diffractive pupil that redistributes light into a highly structured PSF rich in spatial gradients. This pattern creates local astrometric references on the detector, making many optical errors common-mode and improving robustness to systematics such as focus drift and plate-scale changes.

The pupil is realised as a thermally stable, near-achromatic liquid-crystal phase plate with a binary, asymmetric design. By breaking inversion symmetry, it enables direct recovery of both even- and odd-order aberrations from the science images themselves. The pattern was optimised through differentiable forward modelling using Fisher-information metrics to maximise astrometric sensitivity under realistic aberration conditions.

A second innovation incorporates orthogonal diffraction gratings within the pupil, generating four spectral sidelobes that provide simultaneous spectroscopic metrology. Careful phase engineering preserves a complex central PSF for astrometry while producing diffraction-limited sidelobes for wavelength calibration. These measurements disentangle chromatic shifts caused by stellar spectral variations from plate-scale changes induced by focal-length drift, enabling continuous tracking of both effective wavelength and optical scale.

By distributing stellar flux over thousands of pixels, the system suppresses flat-field and intra-pixel sensitivity errors, avoids saturation of bright targets such as alpha Centauri, and turns every science exposure into an in situ wavefront sensor. Combined with pointing dithers and global forward modelling, this architecture continuously constrains detector geometry, pixel gains and optical alignment, delivering the precision required for sub-microarcsecond astrometry.

\section{TOLIMAN Spacecraft and Status} \label{sec:asgard}

The TOLIMAN payload consists of a 12.5 cm Ritchey–Chr\'{e}tien telescope designed for a three-year astrometric survey of alpha~Centauri A and B from low-Earth orbit. The diffractive pupil is mounted at the entrance aperture, together with a 530–640\,nm bandpass (selected to avoid variable chromospheric features), producing the structured point-spread function recorded by a scientific CMOS detector. The telescope employs silicon-carbide optics and structure to maximise thermal and mechanical stability, and is hosted aboard a 16U CubeSat bus.

Recovery of data capable of reaching our micro-arcsecond precision goal requires exceptional control of pointing and thermal stability. Fine image stabilisation is provided by a piezo-driven tip–tilt platform that isolates the payload from spacecraft jitter, while passive insulation, radiators, heaters, and carefully managed spacecraft attitudes minimise thermally induced changes in plate scale and wavefront error.

Science images are recorded at approximately 10 frames per second, with only the regions containing the stellar PSF and spectral reference features retained and, after lossless compression, queued for downlink. Data analysis is based on a differentiable end-to-end model of the optical system implemented in JAX, enabling simultaneous estimation of astrometric position, wavefront state, and detector behaviour. This digital-twin framework supports both instrument calibration and science extraction, with simulations demonstrating recovery of micro-arcsecond signals under increasingly realistic noise conditions.

For exceptionally bright targets such as alpha~Centauri, a 12.5 cm aperture operating near the photon-noise limit can achieve micro-arcsecond sensitivity within only a few hours of integration. By combining narrow-angle astrometry with an internal diffractive reference and simultaneous spectro-metrology, TOLIMAN seeks to overcome the systematic errors that have historically limited ultra-precise astrometry on small spacecraft. 

At the time of writing, most of the major optical components have been fabricated or are at advanced stages of manufacture. Following final assembly of the science payload, integration and testing are slated to be complete for a launch window that opens in mid-2027. The mission serves as a technology demonstrator for future exoplanet-focused astrometry missions, including concepts such as SHERA and THEIA, and advances the broader application of coded-aperture techniques for precision astronomical measurement.

\section{Beyond TOLIMAN: astrometry for isolated stars with fragmented pupils} \label{sec:beyond}

Narrow-angle binary-star astrometric measurement, as embodied in the TOLIMAN mission, enables telescopes of relatively modest aperture to circumvent the fundamental difficulty in getting to micro-arcsecond precision. 
However only a handful of stars -- close binaries with modest contrast ratio -- are suited to this technique. 
This section introduces a conceptual optical system that enables isolated stars to be studied, opening a large population of potential planet-hosting systems.
In particular, the discussion is motivated by bright nearby sun-like single star systems such as Epsilon Eridani or Tau~Ceti.

For telescopes of modest aperture, accumulating sufficient photons\cite{Guyon2012} to establish the astrometric reference frame using available field stars over $\sim$arcminute fields is challenging, requiring optical and electronic systems to deliver accurate registration of thousands of faint objects. 
However if we expand our search to wider scale, $\sim$few-degree fields, statistically there should exist a small handful of bright stars that yield photon counts comparable to the bright target of interest themselves. 
Indeed, for example, this is found to be the case for fields of radius 2.6 degrees surrounding Epsilon Eridani (3.7\,mag) and Tau~Ceti (3.5\,mag) which contain 6 and 4 stars brighter than 6.5\,mag.
The combined photon flux from these field stars is, respectively, 170\% and 54\% of that from the central target star.
However the requirement to image fields spanning several degrees at the diffraction limit places extreme demands on any orthodox optical design, requiring gigapixel sensors and many-element optical trains that are difficult to stabilise.

\begin{figure}[ht] 
    \begin{center}
    \includegraphics[scale=0.5]{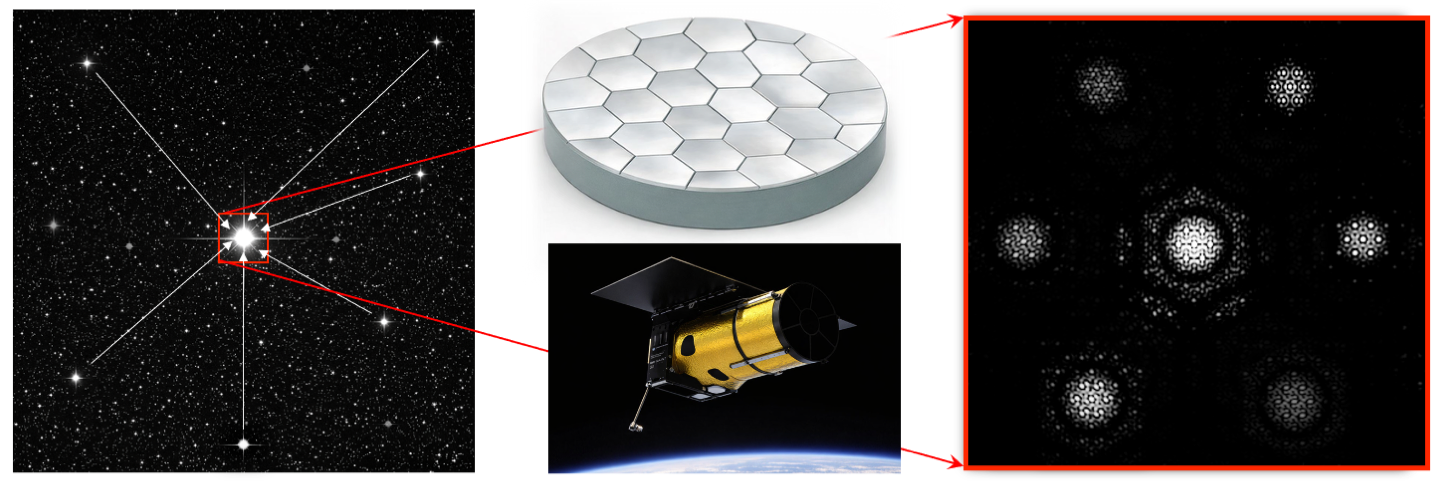}
    \end{center}
    \caption{Schematic diagram of the {\it Fragmented Pupil} telescope concept. Left Panel: Bright stars available over $\sim$degree scale optical fields are condensed into an artificial cluster at sub-arcminute scales (red box). Middle Panel: This is accomplished in a single reflection from a high-stability tessellated optical surface aboard a stable space telescope. Right Panel: Precision astrometric monitoring can then be accomplished by analysis of the resulting well-sampled interferograms in the image plane.
    Here the central science target is surrounded by a set of 6 reference stars, condensed to lie at arcsecond scales.}
    \label{fig:asgard}
\end{figure}

Here we propose instead an unorthodox optical system capable of imaging a sparse subset of widely distributed fields at the diffraction limit, utilising only a modest number of pixels.
This can be accomplished by fabricating a bespoke mirror figure comprising a system that we term a {\em fragmented pupil} telescope.
The purpose is to condense a set of widely distributed, high value bright stars into a concentrated virtual field using only a single, highly stable, optical surface.
Where $n$ stars are to be observed (one science target and $n - 1$ reference stars), the primary mirror is tessellated into $n$ multi-element patterns where each is assigned to a separate star -- hence a separate pointing origin.
After only a single reflection from a high stability optical component, in the image plane the science target is now surrounded by a dense artificial cluster of comparably bright reference stars with which to perform high precision astrometric registration. 

For any single star, due to the sparse nature of the pupil viewing that sightline, the sensor will record an interferogram rather than a simple Airy disc, and furthermore only a scaled fraction of the full collecting area of the mirror is assigned to each object.
Perhaps the most serious limitation of this proposed architecture is that the geometry of the field of desired reference stars must be carved into the optics.
Therefore a bespoke optical system must be designed for every star and it cannot be used on other targets.
Work arounds and alternatives to these limitations are under active study at the time of writing.
Detailed simulations and signal-to-noise calculations will appear in a forthcoming paper.

\acknowledgments 
 
The original scientific impetus as well as critical financial and logistical support for this research
program were provided by the Breakthrough Watch program as part of the Breakthrough Initiatives.
The work builds upon an earlier feasibility study jointly funded by the Breakthrough Initiatives
and The University of Sydney, with further resources from the Australian Research Council grant
LP210200594.
This work was co-funded by the EU and supported by the Czech Ministry of Education, Youth and Sports through the project CZ.02.01.01/00/22\_008/0004596 (SenDISo).
The authors acknowledge the Gadigal people of the Eora Nation upon whose ancestral lands
The University of Sydney now stands. We pay our respects to Elders past and present, and extend
that respect to all Aboriginal and Torres Strait Islander Australians. 

\bibliography{references} 
\bibliographystyle{spiebib} 

\end{document}